\documentstyle[prl,floats,aps,epsfig]{revtex}

\begin{document}

\title{Electrical conductivity of inhomogeneous two component media in
two dimensions.}
\author{V.V. Kabanov, K. Zagar, D. Mihailovic}
\address{Institut Jozef Stefan, Jamova 39, 1000 Ljubljana, Slovenia}
\date{\today}
\maketitle

\begin{abstract}
The electrical conductivity is calculated for regular
inhomogeneous two component isotropic medium in which droplets of
one phase with conductivity $\sigma _{2}$ are embedded in another,
with conductivity $\sigma _{1}$. An expression is formulated which
can be used in many different situations, and is of particular
relevance in the case where the relative proportion of the
components is temperature dependent and varies over a wide range.
The behavior of effective conductivity depends on the spatial
arrangements and the shape of the inclusions.
\end{abstract}


\section{Introduction}

The determination of the effective conductivity $\sigma _{eff}$ of
spatially inhomogeneous heterophase systems is an old, yet
increasingly important problem of theoretical physics. With the
advent of new nanoscale probes of condensed matter systems, it has
been apparent that many very diverse systems which were previously
thought to be homogeneous are in fact either statically or
dynamically inhomogeneous. The effective conductivity in such
cases cannot be dealt with in terms of homogeneous medium theory,
is not trivial, and solutions are known so far only in some rather
special cases. Different aspects of the theory and different
limiting cases are extensively discussed in the Ref.\cite{berstr}.

In this paper we focus on the problem of calculating the effective
conductivity of an inhomogeneous two-dimensional (2D) plane. The classical
problem could be formulated as follows. Let us assume a $2D$ system which
contains a mixture of $N~\ (N\geq 2)$ different phases or materials with
different conductivities $\sigma _{i}$, $i=1,2,...,N$. The arrangements of
different phases could be random or regular. The question which we wish to
address is how the effective conductivity of the plane depends on the
conductivities of the phases, their concentration and the spatial
arrangements of the phases.

In the past, a number of different approaches have been used to tackle this
problem. The exact result for the effective conductivity of a two component
system with a symmetric and isotropic distribution of components was
obtained by Dykhne\cite{dihne}. He found that effective conductivity of the
system is determined by the simple relation $\sigma _{eff}=\sqrt{\sigma
_{1}\sigma _{2}}$. A "symmetric" distribution, applied to this problem is
one in which the two components can be interchanged without changing the end
result. Obviously, one requirement for a symmetric distribution is that the
two components have equal proportions, but it also means that more general
cases cannot be considered with this model.

Further investigations have shown that more general duality
relation is valid for 2D heterogeneous conductors then that
initially considered by Keller and Dykhne\cite{berstr}. More
recently it was shown that more general relation for the tensor of
the effective conductivity exists which is valid for
multicomponent and anisotropic systems\cite{fel,marikhin}. The
effective conductivity of several examples of \emph{ordered}
two-component systems was also calculated
exactly\cite{ovchdug,ovluk,ovch}. It was shown\cite{ovchdug,ovluk}
that for a chess-board plane and for a plane constructed of
triangles, the relation derived by Dykhne is also valid.

A similar relation to the Dykhne formula for the effective conductivity of a
system consisting of randomly distributed metallic and dielectric regions
near a metal-to-insulator transition was derived by Efros and Shklovskii
\cite{efros}. They generalized the expression of Dykhne on the basis of
scaling arguments to the case of arbitrary concentrations of the two phases
near a percolation threshold, so that the effective conductivity becomes:
\begin{equation}
\sigma _{eff}=\sigma _{1}(\sigma _{2}/\sigma _{1})^{s},
\end{equation}
where $s$ is a universal scaling exponent. Critical exponents are
relatively well known as well for this type of system\cite{balag}.
This relation is not applicable when the system is driven away
from the percolation threshold and the general solution of the
effective conductivity of an inhomogeneous medium thus remains an
open problem.

Sen and Torquato \cite{torq89} derived an expression which allows for explicit
calculation of the effective conductivity tensor from $n$-point probability
functions $S_n({\bf r}_1, \dots, \vec{\bf r}_n)$. These functions give the
probability of points at ${\bf r}_1, \dots, {\bf r}_n$ belonging to the same
phase, and are thus uniquely determined by spatial distribution of the phases.
Unfortunately, the application of this method is limited as the computations
involving $n>5$ are fairly time consuming.

Different expansions of the effective conductivity in terms of
small parameter have been used in the past
\cite{berstr,brown,bergman}. In most of the cases the low order
terms weakly depend on microgeometry.  A diagrammatic expansion
for the effective conductivity was developed by Khalatnikov and
Kamenshchik \cite{halat} which promises to give more generally
applicable results. The perturbative approach seems to be quite
effective since it allows us to analyze random and nonsymmetric
distributions with different conductivity.

The problem was discussed for the case $N=2$ and $N=3$ on the
basis of numerical calculations \cite{fel,torq} as well. It was
shown that the effective conductivity for $N=3$ is not universal
and depends on the spatial arrangements of the phases. We have
employed boundary element method for efficient numerical treatment
of two-dimensional multi-phase systems with arbitrary arrangement
of phases. More details on the method and its results can be found
in \cite{zagar}.

In this paper we consider the conductivity of a two-phase system in two
dimensions for a wide range of concentration and conductivities. One phase
is assumed to be composed of droplets (of different shapes) with
conductivity $\sigma _{2}$ embedded within a medium of conductivity
$\sigma_{1}$ (see Fig. 1). We begin by calculating the effective conductivity
$\sigma _{eff}$ using a perturbation theory approach with the two phases
having volume fractions $(1-\nu )$ and $\nu $ respectively. Since the
problem is linear, we can introduce a dimensionless conductivity $\sigma $,
measured in units of $\sigma _{1}=1,$ and the effective conductivity
$\sigma_{eff}$ is a function of $\sigma =\sigma _{2}/\sigma _{1}=\sigma _{2}$
and $\nu$.  The volume-averaged conductivity
$\bar{\sigma}=\frac{1}{V}\int \sigma dV$ is given by:
\begin{equation}
\bar{\sigma}=(1-\nu )+\nu \sigma
\end{equation}

Assuming that the conductivity of the two phases is not vastly different
$|\sigma -1|\ll 1$, the effective conductivity could be calculated by
perturbation theory\cite{halat}. To apply perturbation theory we rewrite the
spatial dependence of conductivity as:
\begin{equation}
\sigma({\mathbf r})=\bar{\sigma}(1-\alpha({\mathbf r})),
\end{equation}
where $\alpha(\mathbf{r})=\frac{\sigma(\mathbf{r})-\bar{\sigma}}{\bar{\sigma}
}$. Then, assuming that the spatial distribution of conductivity is
uncorrelated, we obtain:
\begin{equation}
\int d{\mathbf r}\alpha({\mathbf r})\alpha({\mathbf r+r^{\prime}}
)=\frac{(\sigma-1)^{2}\nu(1-\nu)}{\bar{\sigma}^{2}}
\delta({\mathbf r^{\prime}})
\end{equation}
After a straightforward calculation, the conductivity up to second order in
$\alpha $ is given by:
\begin{equation}
\sigma _{eff}=\bar{\sigma}-\frac{(\sigma -1)^{2}\nu (1-\nu )}{2\bar{\sigma}}
\end{equation}
This result has been known for many years and was derived for the
dielectric function of dielectric mixtures\cite{landau}. In Refs.
\cite{brown,bergman} it was also derived using a systematic
perturbation expansion which showed it to be exact to second order
in $\alpha$. The second term in Eq.(5) represents the first
non-vanishing contribution due to the inhomogeneity of the
distribution of the phases. For the case $\nu=0.5$, the result
coinsides with the expansion of the exact expression for the
conductivity up to the second order in $(\sigma_{2}-\sigma_{1})$
\cite{dihne}:
\begin{equation}
\sigma _{eff}=\sqrt{\sigma _{1}\sigma _{2}}.
\end{equation}

\section{Conductivity of a regular isotropic two-component system in 2D.}

Next, we calculate exactly the effective conductivity of the plane with
different regular isotropic distributions.
As before, let us consider a 2D plane which
is constructed from two different phases with different conductivities
$\sigma _{1}=1$ and $\sigma _{2}=\sigma $. The regions with conductivity
$\sigma _{2}$ have a circular shape with radius $R$ and form a regular square
lattice with the period $a$ as shown in Fig.1a. Changing the radius $R$ from
0 to $R=a/2$ we can change the volume fraction of the second phase from
$\nu =0$ to a critical concentration $\nu _{c}=0.785$ whereafter the regions
with conductivity $\sigma _{2}$ start to overlap and a percolation threshold
is reached. In the case of metallic droplets, the total charge density
should be zero, while a finite charge density can accumulate on the surface
between different phases. This allows us to formulate the integral equation
for the surface charge density\cite{ovchdug,ovluk}. Let us define a surface
charge density by the relation $\rho (\theta )Rd\theta =d\rho (\theta )$,
where $d\rho (\theta )$ is the charge on the small part of the surface
between the two components with the length $dl=Rd\theta $. Taking into
account that the scalar potential at the point $\mathbf{r}$ is determined by
the relation
\begin{equation}
\phi =E_{0}x-2\int d^{2}r^{^{\prime }}\ln {|\mathbf{r}-\mathbf{r}^{^{\prime
}}|}\rho (\mathbf{r}^{^{\prime }})
\end{equation}%
where $\ln {|\mathbf{r}-\mathbf{r}^{^{\prime }}|}/2\pi $ is the 2D
Green's function. The boundary conditions on the surface between
two phases are \cite{landau}:
\begin{equation}
E_{n}^{1}-E_{n}^{2}=4\pi \rho (\theta )
\end{equation}%
\begin{equation}
\sigma _{1}E_{n}^{1}=\sigma _{2}E_{n}^{2}
\end{equation}
Substituting ${\bf r^{'}} = (ma+R\cos{(\theta^{'})}){\bf
i}+(na+R\sin{(\theta^{'})}){\bf j}$ and ${\bf r} =
R\cos{(\theta)}{\bf i}+R\sin{(\theta)}{\bf j}$ to Eqs.(7-9) we
obtain integral equation for the surface charge density in the
following form:
\begin{eqnarray}
\rho(\theta)=\frac{\kappa}{2\pi}\Bigl[ E_{0}\cos{(\theta)} +
2r\sum_{n,m=-\infty}^{\infty}\int_{-\pi}^{\pi}
d\theta^{'}\rho(\theta^{'}) \nonumber \\
\Re\Bigl (\frac{\exp{(i\theta)}}
{m+r(\cos{(\theta^{'})}-\cos{(\theta)})+
i(n+r(\sin{(\theta^{'})}-\sin{(\theta)}))}\Bigr )\Bigr ]
\end{eqnarray}
where $r=R/a$, and $\kappa =\frac{(1-\sigma )}{(1+\sigma )}$. As
it is shown in the Appendix the sum over $m$ can be calculated
exactly and the integral equation for surface charge density will
be reduced to the following form:
\begin{equation}
\rho (\theta )=\frac{\kappa }{2\pi }[E_{0}\cos {(\theta )}+2r\sum_{n=-\infty
}^{\infty }\int_{-\pi }^{\pi }d\theta ^{^{\prime }}K(n,\theta ,\theta
^{^{\prime }})\rho (\theta ^{^{\prime }})]
\end{equation}
where
\begin{equation}
K(n,\theta ,\theta ^{^{\prime }})=\pi\frac{\cos{(\theta )}\sin
{(2\pi r(\cos { (\theta ^{^{\prime }})}-\cos {(\theta )}))}+\sin
{(\theta )}\sinh {(2\pi (n+r(\sin {(\theta ^{^{\prime }})}-\sin
{(\theta )})))}}{\cosh {(2\pi (n+r(\sin {(\theta ^{^{\prime
}})}-\sin {(\theta )})))}-\cos {(2\pi r(\cos { (\theta ^{^{\prime
}})}-\cos {(\theta )}))}}
\end{equation}

Expanding the surface density $\rho (\theta )$ in terms of Legendre
polynomials $P_{l}(\cos {(\theta )}),$ and taking into account that $\rho
(-\theta )=\rho (\theta )$ and $\rho (\pi -\theta )=-\rho (\theta )$:
\begin{equation}
\rho (\theta )=\sum_{l=1}^{\infty }c_{2l-1}P_{2l-1}(\cos {(\theta )})
\end{equation}
we obtain the following linear set of algebraic equations for the
coefficients $c_{2l-1}$
\begin{equation}
\frac{2c_{2l-1}}{4l-1}=\frac{\kappa }{2\pi
}[\frac{2}{3}E_{0}\delta _{l,1}+2r\sum_{k=1}^{\infty
}c_{2k-1}K_{l,k}]
\end{equation}
where
\begin{equation}
K_{l,k}=\sum_{n=-\infty }^{\infty }\int_{0}^{\pi }d\theta ^{^{\prime
}}\int_{0}^{\pi }d\theta K(n,\theta ,\theta ^{^{\prime }})\sin {\theta }
P_{2l-1}(\cos {(\theta )})P_{2k-1}(\cos {(\theta ^{^{\prime }})})
\end{equation}

Solving Eq.(14) taking into account a finite number of Legendre
polynomials we obtain surface charge density Eq.(13). As a result
the effective conductivity is evaluated by calculating the total
current $j=\sigma_{1}E_{n}=E_{n}$ through the semicircular surface
with the radius $R^{^{\prime }}=a/2$ (see Fig.1a). Calculations,
similar to that of Eq.(12), lead to the expression for effective
conductivity:
\begin{equation}
\sigma _{eff}=\frac{\kappa }{4\pi }\int_{-\pi /2}^{\pi /2}d\theta \lbrack
\cos {(\theta )}+\frac{2r}{E_{0}}\sum_{n=-\infty }^{\infty }\int_{-\pi
}^{\pi }d\theta ^{^{\prime }}K^{^{\prime }}(n,\theta ,\theta ^{^{\prime
}})\rho (\theta ^{^{\prime }})]
\end{equation}
where
\begin{equation}
K^{^{\prime }}(n,\theta ,\theta ^{^{\prime }})=\pi\frac{\cos
{(\theta )}\sin { (2\pi (r\cos {(\theta ^{^{\prime }})}-\cos
{(\theta )}/2))}+\sin {(\theta )} \sinh {(2\pi (n+r\sin {(\theta
^{^{\prime }})}-\sin {(\theta )}/2))}}{\cosh { (2\pi (n+r\sin
{(\theta ^{^{\prime }})}-\sin {(\theta )}/2))}-\cos {(2\pi (r\cos
{(\theta ^{^{\prime }})}-\cos {(\theta )}/2))}}
\end{equation}

The result above applies to the case of a uniform distribution of
circular droplets within the plane. To see how the effective
conductivity depends on the shape of the regions with conductivity
$\sigma _{2},$ we have performed calculations for the case where
of circular droplets we substituted with squares, triangles and
rhombuses with the ratio of diagonals $\tan{\alpha}=a/b$ where $a$
and $b$ translation vectors along $x$ and $y$ respectively (see
Fig.1b,c,d). In all these cases Eqs.(10-17) are slightly modified
since in polar coordinate system $r(\theta)$ is a function of
angle. Contrary to the case of circles, percolation threshold for
cases b,c, and d is $\nu _{c}=0.5$. Note that in the case of
rhombuses the lattice is anisotropic and $\sigma_{eff}^{11} \ne
\sigma_{eff}^{22}$.

\section{Discussion.}

The results of the calculations of the effective conductivity are
presented in Fig.2 as a function of $\sigma^{\nu}$ for different
values of volume fraction $\nu$. It is easy to check that the
results satisfy the generalized duality
relation\cite{fel,marikhin}:
\begin{equation}
\sigma_{eff}^{11}(\sigma_{1},\sigma_{2})\sigma_{eff}^{22}(1/\sigma_{1}
,1/\sigma _{2})=1.
\end{equation}
% ...
For the cases of circles, squares and triangles
$\sigma^{22}=\sigma^{11}$. In the case of rhombuses
$\sigma^{22}(\alpha)=\sigma^{11}(\pi/2-\alpha)$.
 Fig.2 (a,b,c,d) shows that for small $\kappa$, perturbation
theory\cite{brown,bergman,halat} (Eq.5) gives the correct result
independent of geometry.

\subsection{Approximate expression for the effective conductivity.}

Although the predictions in Fig.2 represent the results of a
precise numerical calculation, they are not very tractable when it
comes to comparing with experimental data, being the result of
numerical calculations. It is therefore helpful to try and obtain
a functional form for describing the behavior predicted in Figure
2, which also includes all the relevant parameters, such as volume
fraction $\nu$, and the two conductivities $\sigma_{1}$ and
$\sigma_{2}$. Such an expression can then be used for a wide range
of problems, provided the range of validity is taken into account.
We describe the properties of such a heuristically determined
function and determine its range of validity in terms of
parameters $\nu,\sigma_{1}$ and $\sigma_{2}$.

As can be seen from the Figure 2, the dependence of the effective
conductivity on $\sigma$ shows similar behavior, independent on
the particular geometry of the phases. First, we observe that when
$\kappa$ is small all the curves are linear in $\sigma^{\nu}$ with
the same slope. In the relatively wide interval of sigma
($0.1<\sigma<10$) the effective conductivity is determined by the
equation:
\begin{equation}
\sigma_{eff}(\sigma )=\sigma _{1}^{(1-\nu )}\sigma _{2}^{\nu}.
\end{equation}
The range of applicability of this formula becomes wider as we
approach the percolation threshold $\nu_{c}$. When
$\sigma=\sigma_{2}/\sigma_{1} \gg 1$ effective conductivity
saturates at $\sigma_{sat}$. $\sigma_{sat}$ is not universal and
depends on the geometry. Recently it was pointed out that for the
case of circles for $\nu <0.5$ in the whole range of $\sigma$
effective conductivity may be approximated by the
formula\cite{emets}:
\begin{equation}
\sigma_{eff}(\kappa )=(1-\nu\kappa)/(1+\nu\kappa).
\end{equation}
To derive approximate expression for effective conductivity we
assume that Eq.(20) remains correct if we substitute instead of
$\nu$ the effective volume fraction $\nu_{eff}(\kappa,\nu)$. We
require that $\nu_{eff}(\kappa,\nu) \approx \nu$ for $\kappa \to
0$ or $\nu \to 0$, and $\nu_{eff}(\kappa,\nu) \approx
\frac{1}{\kappa}\frac{1-\sigma^{\nu_{c}}}{1+\sigma^{\nu_{c}}}$ for
$\nu \to \nu_{c}$ to satisfy Eq.(19) which is valid at
$\nu=\nu_{c}$. It is easy to see that the function:
\begin{equation}
\nu_{eff}(\kappa,\nu) =\nu + \frac{1}{\kappa}\frac{1-\Bigl (
\frac{1-(1-p(\nu))\kappa}{1+(1-p(\nu))\kappa}\Bigr )^{\nu_{c}}}
{1+\Bigl ( \frac{1-(1-p(\nu))\kappa}{1+(1-p(\nu))\kappa}\Bigr )
^{\nu_{c}}}-(1-p(\nu))\nu_{c},
\end{equation}
where $p(\nu) \to 0$ as $\nu \to \nu_{c}$ and $p(\nu) \to 1$ as
$\nu \to 0$, satisfies all of the above requirements. Function
$p(\nu)$ is not universal and depends on the geometric shape of
the region with conductivity $\sigma$ and on the particular
arrangement of these inclusions in the 2D plane. In Fig.3 we plot
$p(\nu)$ as function of $1-\nu/\nu_{c}$ for the cases a,b,and c
respectively. The case d is different, since the effective
conductivity is anisotropic. As it is clearly seen from Figure 3
the behavior of the function $p(\nu)$ for circles (case a) is
different from the cases of squares and triangles (b,c). On the
other hand in the cases b and c $p(\nu)$ shows similar behavior.

\subsection{Shape dependence of the effective conductivity.}

The function $p(\nu)$ is connected with the value of
$\sigma_{sat}=(1+\nu_{eff}(\kappa=-1,\nu))/(1-\nu_{eff}(\kappa=-1,\nu))$.
Therefore the behavior of the function $p(\nu)$ close to
percolation threshold should be different for different
geometries. In Fig.4 we plot the value of $\sigma_{sat}$ as a
function of $(1-\nu/\nu_{c})$ for the case of circles, squares and
triangles. There is an important difference between these two
cases. In the case of circles $\sigma_{sat}$ diverges as a power
of $(1-\nu/\nu_{c})^{-k}$ ($k \approx 0.5$). For the case of
squares and triangles this behavior is logarithmic. In both cases
close to percolation threshold $\sigma_{sat}$ is proportional to
average inverse distance between boundaries of the neighboring
circles or squares $\sigma_{sat} \propto \int dy/(1-2f(y))$ and
$f(y)=\sqrt{(r^2-y^2)}$ for circles and $f(y)=r-|y|$. Here we
assume that the period of the system is 1, and the dimensionless
size of the circle and the square is r. Direct integration leads
to the following results:
\begin{equation}
\sigma_{sat} \propto (\pi/2
-arccos((1-\nu/\nu_{c})^{1/2})/(1-\nu/\nu_{c})^{1/2}-\pi/2
\end{equation}
for the case of circles and
\begin{equation}
\sigma_{sat} \propto -\ln{(1-(\nu/\nu_{c})^{1/2})}
\end{equation}
for the case of squares (Fig.4). For the case of triangles (c) the
asymptotic is similar to Eq.(23) with different numeric
coefficients. Interestingly this observation suggests that
behavior of the function $p(\nu)$ is different depending on the
$curvature$ of the embedded regions.

\section{Conclusion}

From calculations of the effective conductivity of inhomogeneous
two-phase systems in two dimensions we find that the results of
precise numerical calculations can be approximated by a universal
function for $\sigma _{eff}$ Eqs.(20,21), where the function
$p(\nu)$ depends on the spatial arrangements of the 2D plane $and$
on the shape of the inclusions with conductivity $\sigma$. It is
shown that in a large interval of conductivity $\sigma$, the
effective conductivity $\sigma _{eff}$ is determined by the
spatial average of the logarithm of individual conductivities. The
closer the system is to the percolation threshold, the larger the
range of validity of this result. For large values of conductivity
$\sigma$, $\sigma_{eff}$ saturates at the value $\sigma_{sat}$.
The value of $\sigma_{sat}$ near percolation threshold is
determined by the average inverse distance between boundaries of
neighboring regions with conductivity $\sigma$ in the direction of
the field (Eqs.22,23)

The model which we have developed is quite generally applicable and can be
applied in some interesting situations, such as cuprates and other
two-dimensional complex transition metal oxides which exist near a
phase-separation threshold. Importantly, there appears to be a significant
amount of experimental evidence that many anomalous properties of oxides are
associated with the coexistence of two or more phases. The application of the
presented model may help in understanding the transport properties of such
systems.

\section{Appendix}
Here we show how the sum over m in the Eq.(10) could be calculate
exactly. Let us represent the sum in the form:
\[
S=\sum_{m=-\infty }^{\infty}
\Re(\frac{\exp(i\theta)}{(m+\beta+i\alpha)})
\]
where $\beta =r(\cos(\theta^{'})-\cos(\theta))$,
$\alpha=n+r(\sin(\theta^{'})-\sin(\theta))$. Sum over m is
calculated using the definition of di-gamma function. As a result
we obtain expression for the sum:
\[
S=
\Re\Bigl(\exp(i\theta)[\psi(-\beta-i\alpha)-\psi(1+\beta+i\alpha)]\Bigr)=
\pi\Re\Bigl(exp(i\theta)\cot(\pi(\beta+i\alpha))\Bigr)
\]
Calculating imaginary part of previous equation we arrive to the
results Eq.(12):
\[
S=\pi\frac{\cos{(\theta)}\sin{(2\pi\beta)}+\sin{(\theta)}\sinh{(2\pi\alpha)}}
{\cosh{(2\pi\alpha)}-\cos{(2\pi\beta)}}
\]

\vspace{1.5cm}

Figure 1. Spatial arrangements of phases with conductivities
$\sigma_{1}=1$ and $\sigma_{2}=\sigma$ for four considered cases.

Figure 2. Effective conductivity of the plane as a function of
$\sigma^{\nu}$ for different volume fractions and four considered
geometries.

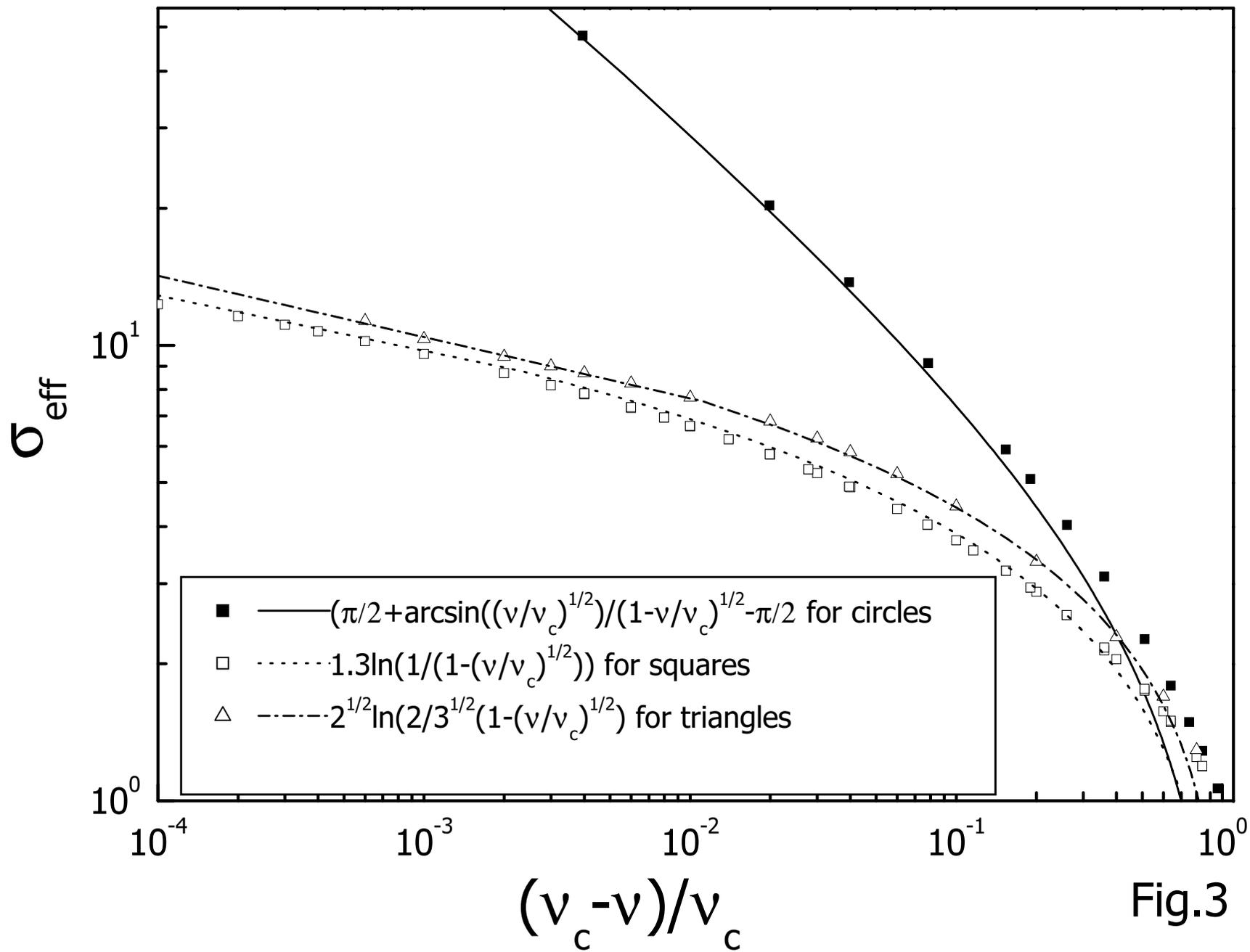
Figure 3. Dependence of the function $p(\nu)$ on $1-\nu/\nu_{c}$
for the cases a,b and c.

Figure 4. Saturated effective conductivity as $\sigma \to \infty$
for the case of a, b and c. Full, dotted and dashed lines show
different analytical asymptotical behavior for these cases.

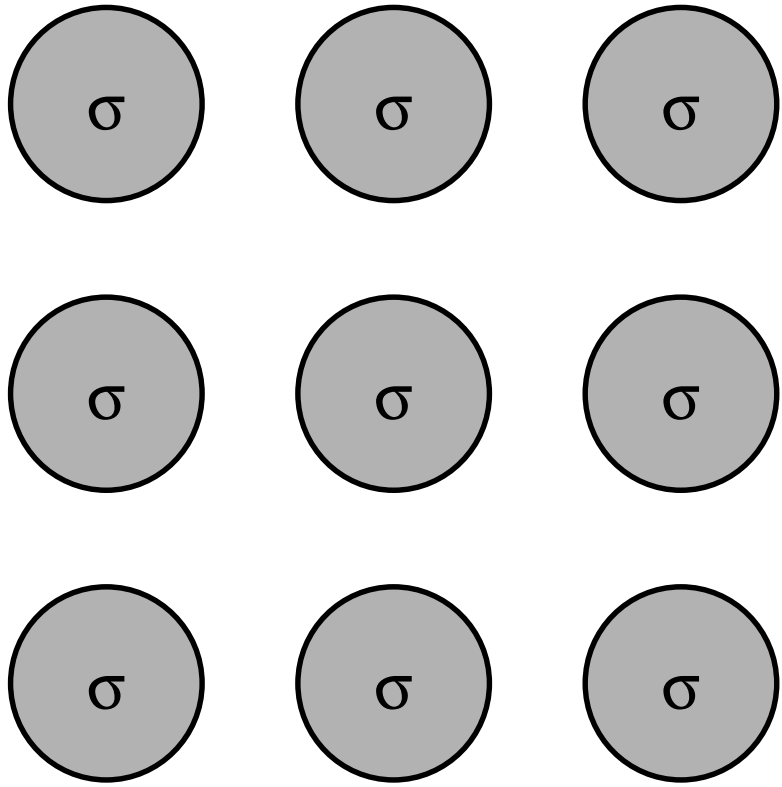
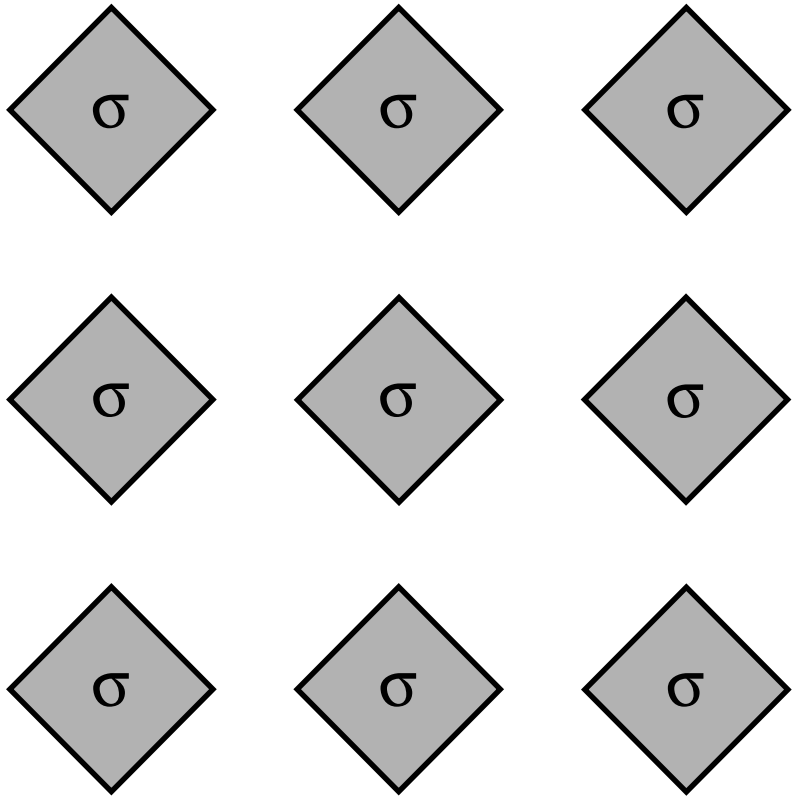
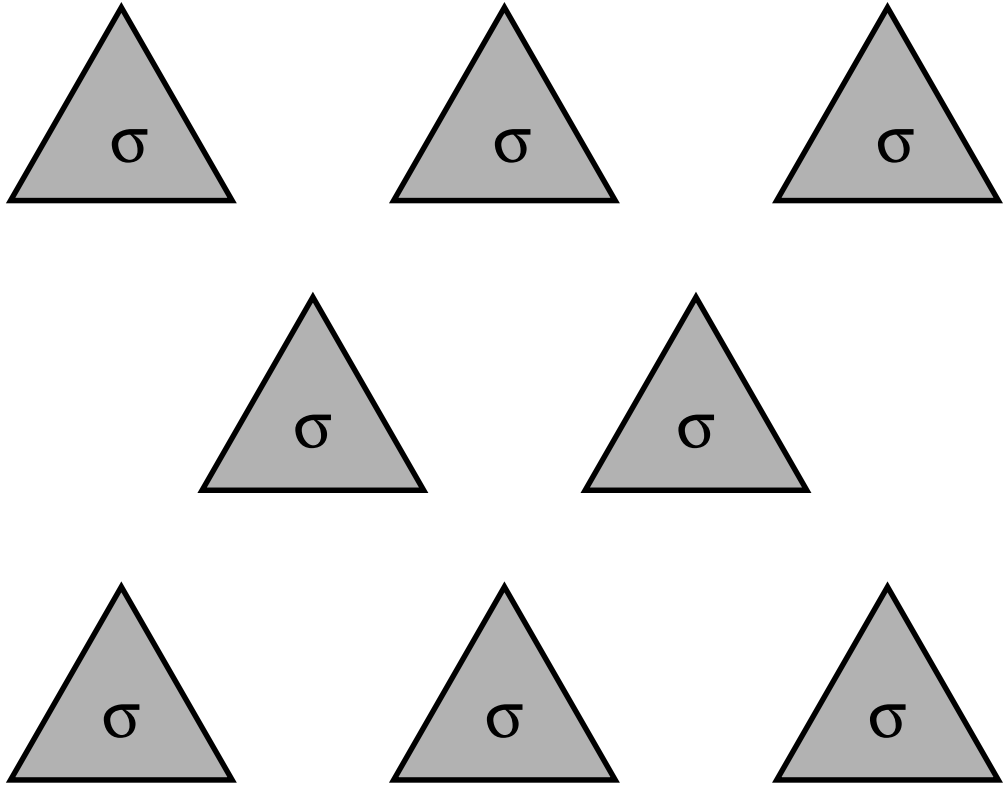
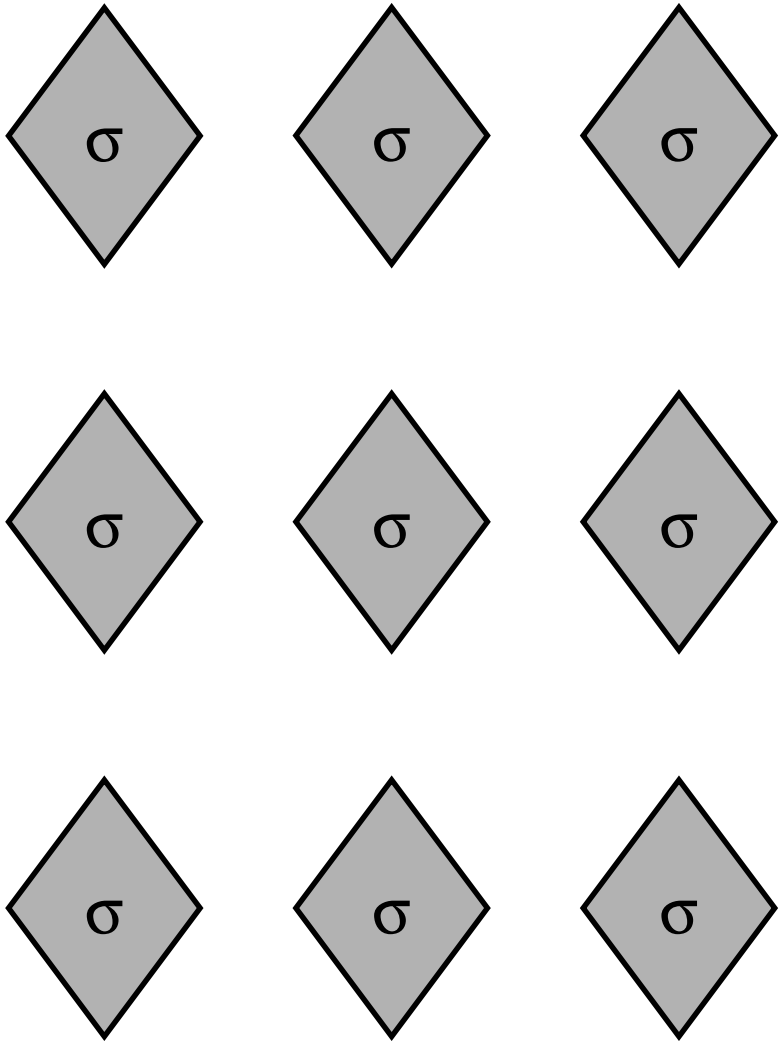
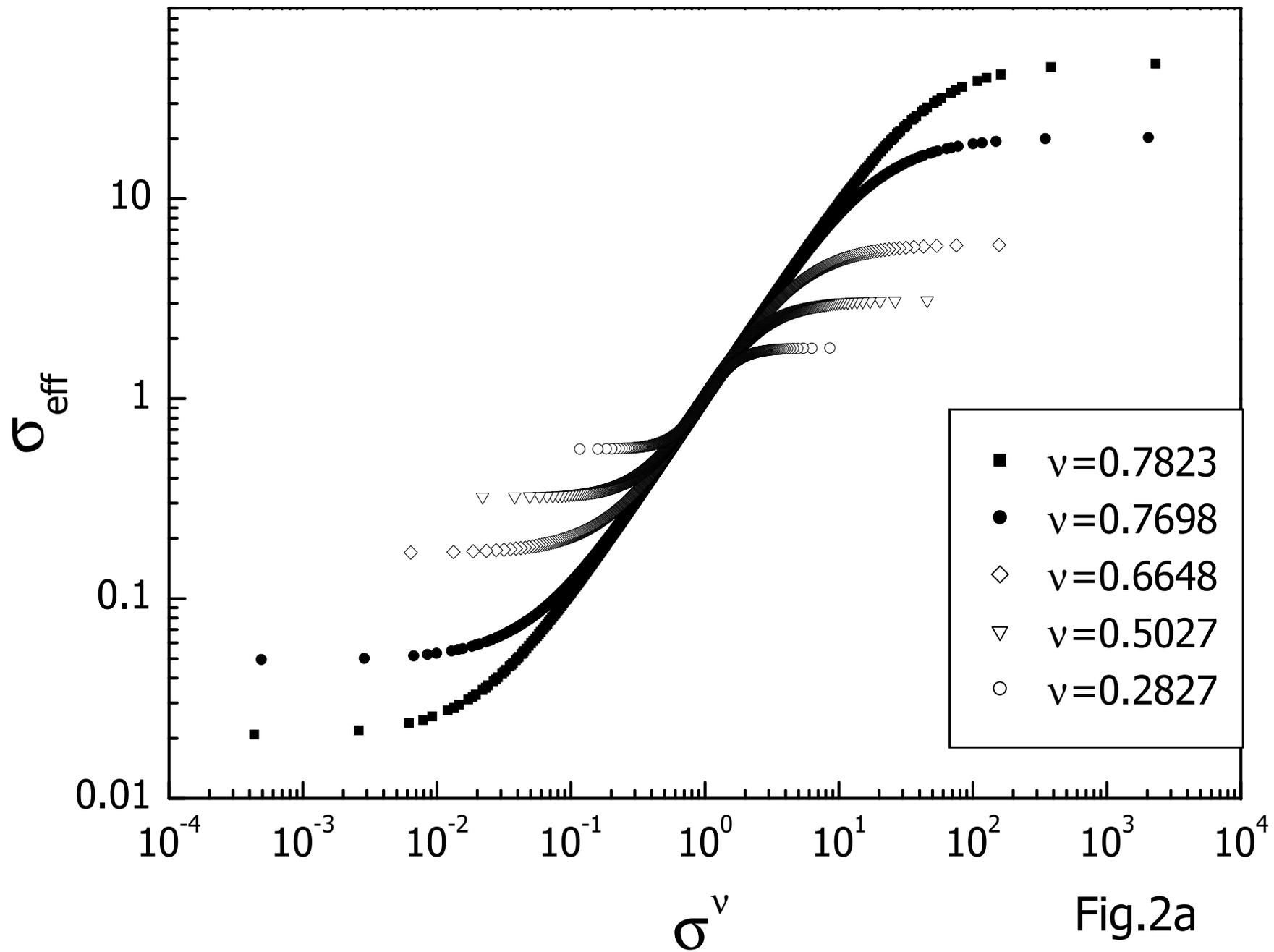
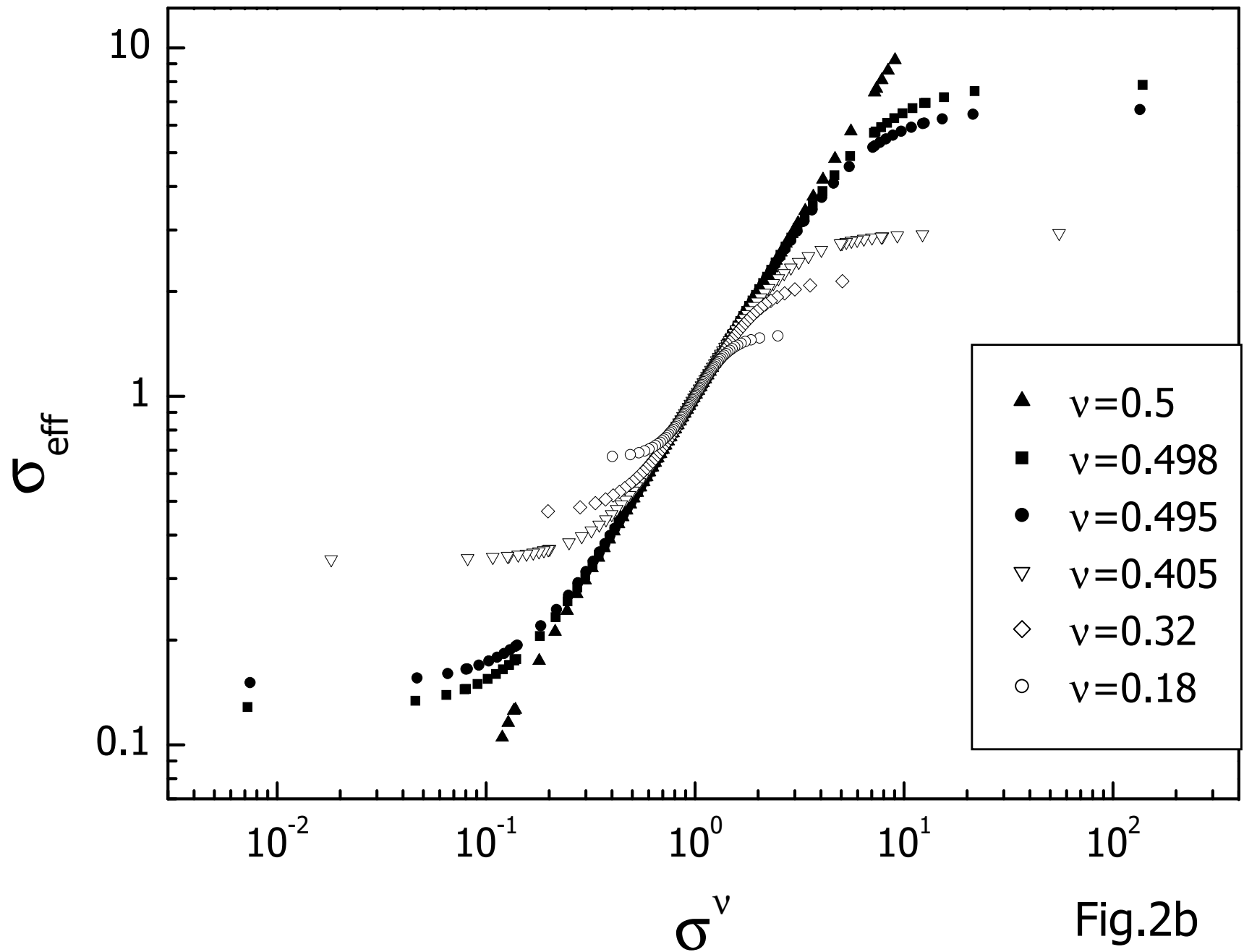
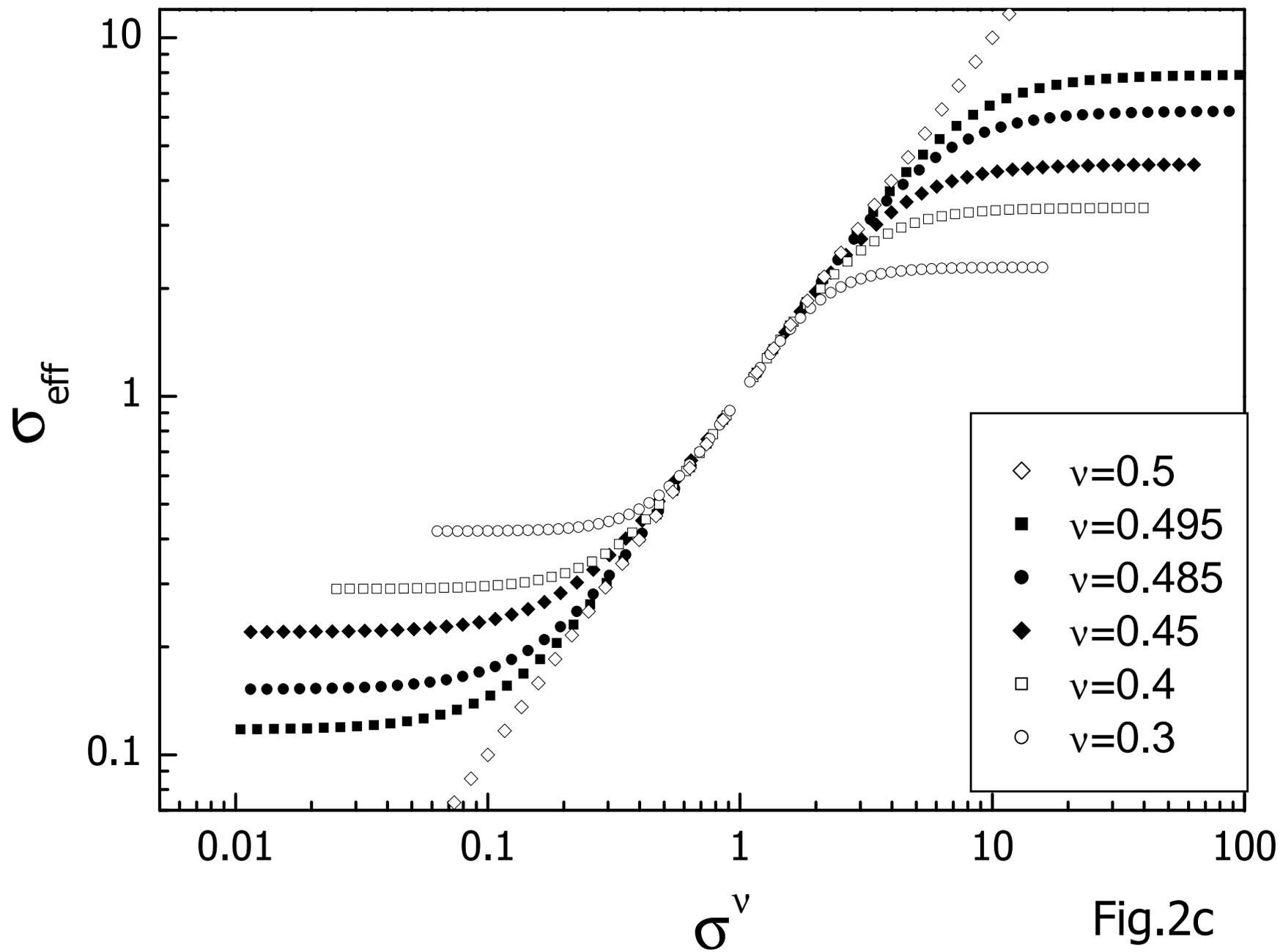
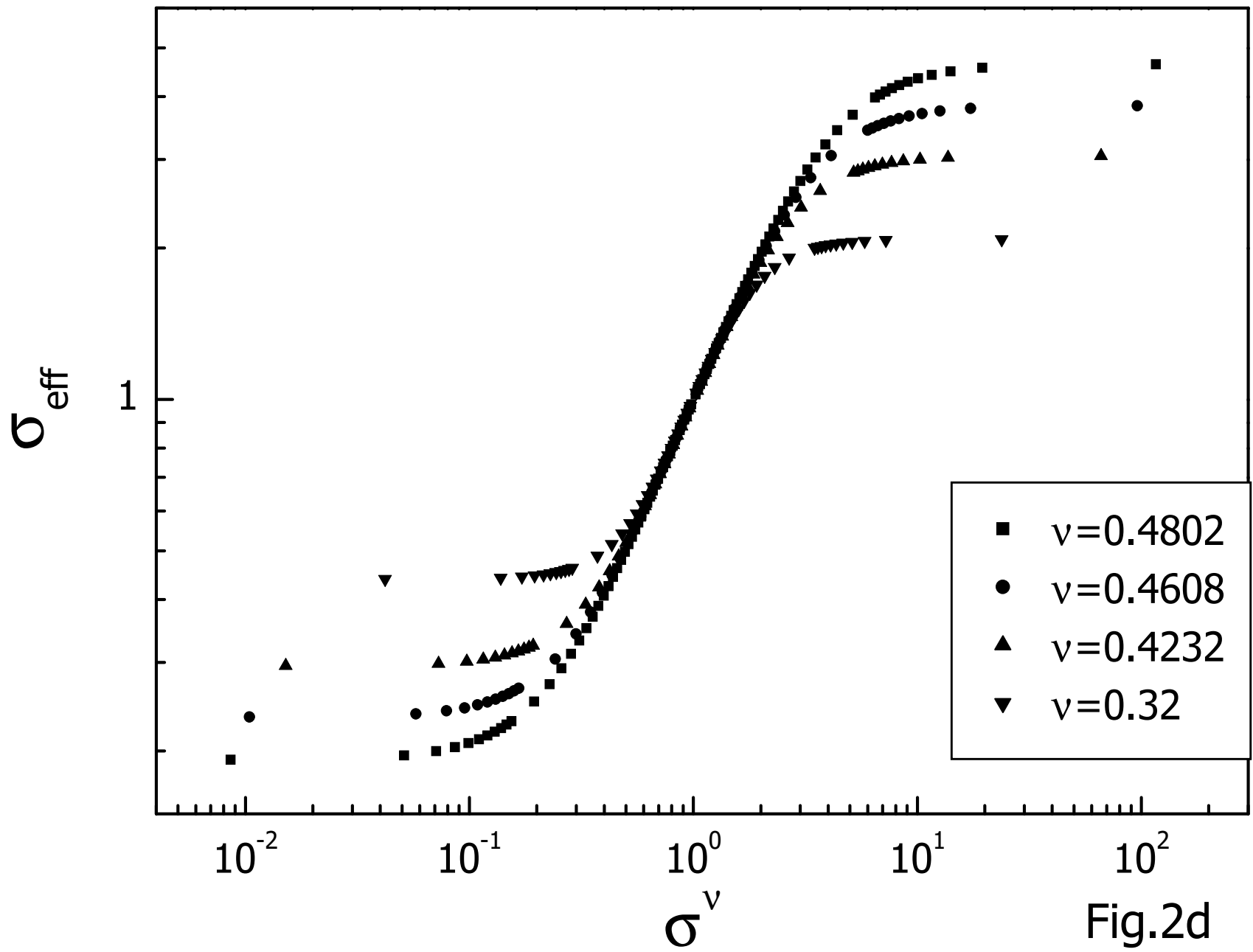
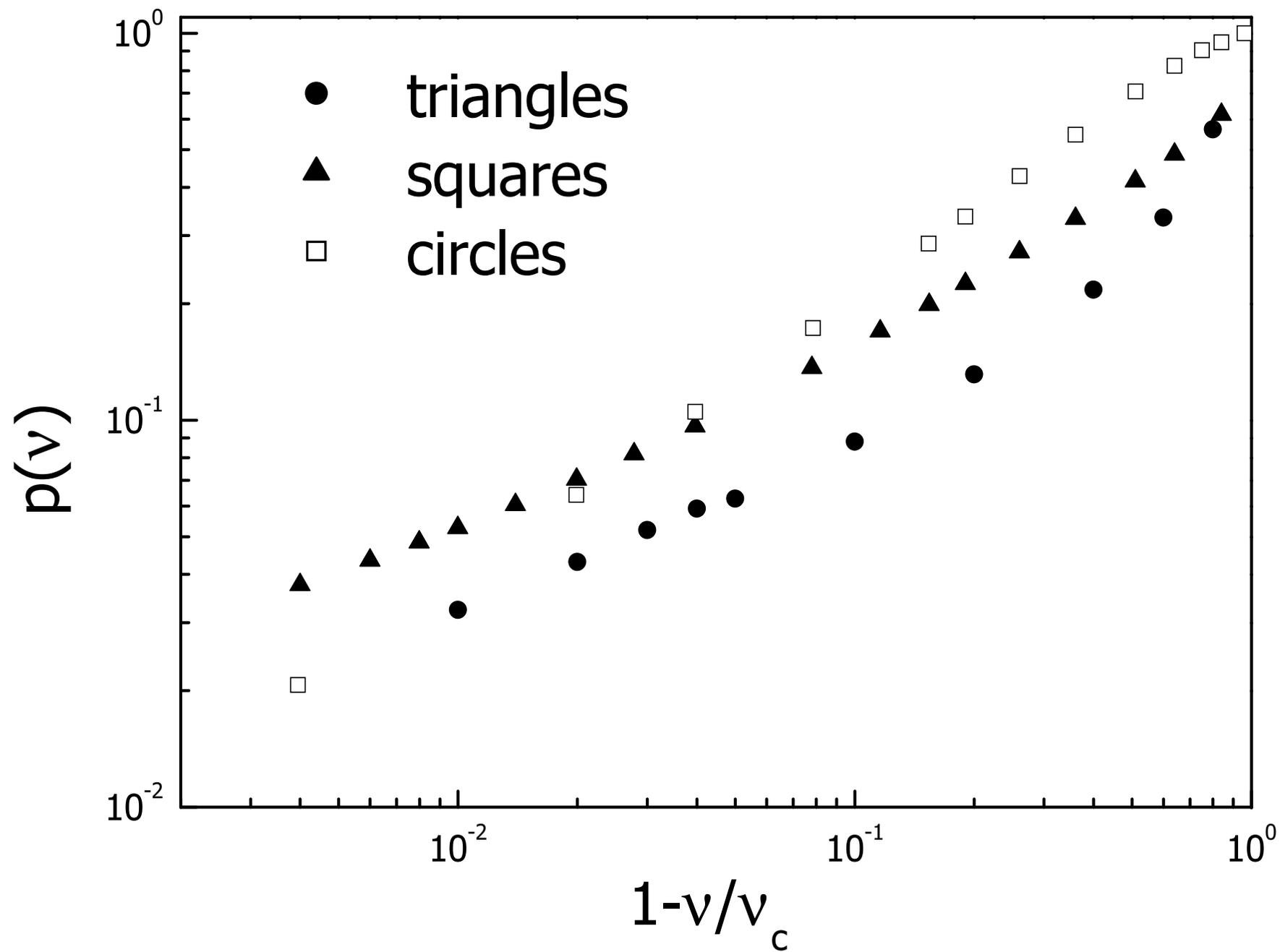
\end{document}